\documentclass[11pt]{article}
\usepackage{epsfig,amsmath}

\title{\bf Axially Symmetric Post-Newtonian Stellar Systems}

\author{
{\sc Camilo Akimushkin}\thanks{e-mail: camilo.akimushkin@gmail.com}
\and
{\sc Javier Ramos-Caro}\thanks{e-mail: javiramos1976@gmail.com}
\and
{\sc Guillermo A. Gonz\'alez}\thanks{e-mail: guillego@uis.edu.co}\\
\newline
\and
Escuela de F\'isica,  Universidad Industrial de Santander\\
A. A. 678, Bucaramanga, Colombia}

\date{ }

\begin{document}

\maketitle

\begin{abstract}
We introduce a method to obtain self-consistent, axially symmetric, thin
disklike stellar models in the first post-Newtonian (1PN) approximation. The
models obtained are fully analytical and corresponds to the post-Newtonian
generalizations of classical ones. By introducing in the field equations
provided by the 1PN approximation a known distribution function (DF)
corresponding to a Newtonian model, two fundamental equations determining the
1PN corrections are obtained, which are solved using the Hunter method. The
rotation curves of the 1PN-corrected models differs from the classical ones and,
for the generalized Kalnajs discs, the 1PN corrections are clearly appreciable
with values of the mass and radius of a typical galaxy. On the other hand, the
relativistic mass correction can be ignored for all models.\\

\end{abstract} 

\section{Introduction}\label{introduccion}

The stars observed in the universe tend to cluster in huge self-gravitating
systems, being the galaxies among the most noticeable and studied of them.
Galaxies can be described by models whose mass is distributed symmetrically
about an axis to a finite distance of it i.e., the galaxy radius. Particularly
useful models are those restricted to a plane or thin disk, since for many
galaxies like the Milky Way its height is small compared to its radius. For
galaxies in general, the average time between collisions or individual meetings
(mean collision time) is greater than the system's life time. Many of the
current models take Newton's law as their field equations \cite{LB, mestel, T1,
T2, HunterToomre, KAL1, jiang00, jiangmos02, GR, jiang, PRG}. However, many
models have been developed recently under general relativity \cite{Lemos1,
Lemos2, Lemos3, gonzalez-letelier, gonzalez-letelier2, semerak-zacek1,
semerak-zacek2, semerak-zacek3}, being one of the main motivations for including
corrections made by general relativity the actual incompatibility between the
rotation curves of theoretical models and the ones observed.

The post-Newtonian approximation introduces general relativity through a series
over the speed $(v/c)$. For Minkowski's background metric, the first
post-Newtonian corrections (1PN) are included taking the terms
\begin{eqnarray}
g_{00} &\approx& -1 + \stackrel{2 \: \: \: \: }{g_{00}} + \stackrel{4 \: \: \:
\: }{g_{00}}, \\
g_{i0} &\approx& \stackrel{3 \: \: \: \: }{g_{i0}}, \\
g_{ij} &\approx& \delta_{ij} + \stackrel{2 \: \: \: \: }{g_{ij}},
\end{eqnarray}
where the upper index denotes the order of the power of $(v/c)$ of the term.
Using harmonic coordinates, the 1PN order potentials are defined as
\begin{eqnarray}\label{ec_pot_pn_def}
\stackrel{2 \: \: \: \: }{g_{00}} &\equiv & -2\phi /c^2, \\
\stackrel{3 \: \: \: \: }{g_{i0}} &\equiv & \zeta_i /c^3, \\
\stackrel{4 \: \: \: \: }{g_{00}} &\equiv & -2(\phi^2 +\psi) /c^4. 
\end{eqnarray}
Henceforth we consider only the stationary case, so the explicit
time-dependent terms along with the potential vector $\boldsymbol{\zeta}$
disappear in all equations.

The stationary field equations in the 1PN approximation are
\begin{eqnarray}
\nabla^2 \phi &=& \frac{4\pi G}{c^2} \stackrel{0 \: \: \: \: }{T^{00}},
\label{ec_campo_pn1} \\
\nabla^2 \psi &=& 4\pi G ( \stackrel{2 \: \: \: \: }{T^{00}} + \stackrel{2 \: \:
\: \: }{T^{ii}} ), \label{ec_campo_pn2}
\end{eqnarray}
where, in the classical limit, potential $\psi$ vanishes and $\phi$ tends to the
Newtonian potential $\phi_N$. Whereas that
\begin{equation}
\frac{\mathrm{d} \mathbf{v}}{\mathrm{d} t} = - \nabla \phi - \frac{1}{c^{2}}
\left[\nabla \left(2 \phi^{2} + \psi \right) + 4 \mathbf{v} (\mathbf{v} \cdot
\nabla \phi) - v^{2} \nabla \phi \right] \label{ecmovimiento1PN}
\end{equation}
is the stationary equation of motion.

A complete statistical description is achieved by knowing the distribution
function (DF) of the system. The DF satisfies a continuity equation in the phase
space called the Boltzmann equation. In the 1PN approximation, the collisionless
Boltzmann equation (CBE) for a many identical particles system is given by
\cite{Rezania}
\begin{equation}\label{ec_general derezania}
v^i \frac{\partial F}{\partial x^i} - \frac{\partial \phi}{\partial x^i}
\frac{\partial F}{\partial v^i}- \frac{1}{c^2} \left( \frac{\partial
\phi}{\partial x^i} (4 \phi + v^2) - \frac{\partial \phi}{\partial x^j} v^i v^j
+ \frac{\partial \psi}{\partial x^i} \right) \frac{\partial F}{\partial v^i} =
0.
\end{equation}
According to Jeans theorem \cite{BT}, the solution of colissionless Boltzmann
equation is any function of the integrals of motion. Now, it is easy to verify
that two isolated integrals a 1PN system with axial symmetry are given by
\begin{equation}\label{ec_ene}
E = \frac{1}{2}v^2 + \Phi,
\end{equation}
where
\begin{equation}
\Phi = \phi + \frac{2\phi^2 + \psi}{c^2},
\end{equation}
and
\begin{equation}\label{lz_pn}
L_z = Rv_{\varphi}e^{-\phi/c^2} \approx Rv_{\varphi}(1-\phi/c^2),
\end{equation}
which can be interpreted as the 1PN generalizations of energy and the $z$
component of angular momentum, respectively.

In addition, the DF must satisfy the condition of self-consistently generating
the macroscopic mean values. In the post-Newtonian approximation, the following
components of the energy-momentum tensor are need,
\begin{eqnarray}
\stackrel{0 \: \: \: \: }{T^{00}}({\bf x},t) &=& c^2\int F({\bf x},{\bf
v},t)d^{3}v, \label{T000} \\
\stackrel{2 \: \: \: \: }{T^{00}} + \stackrel{2 \: \: \: \: }{T^{ii}} &=& 2\int
(v^{2}({\bf x},t) +\phi ({\bf x},t)) F({\bf x},{\bf v},t)d^{3}v. \label{T002}
\end{eqnarray}
Therefore, in the post-Newtonian approximation, self-consistent equilibrium
models are defined by two scalar potentials, $\phi$ and $\psi$, together with a
DF that satisfies 1PN CBE and relations (\ref{T000}) - (\ref{T002}) generating
self-consistenly the 1PN components of the energy-momentum tensor. In this paper
we present a method to obtain post-Newtonian axially symmetric equilibrium
models. The method uses thin disk models, allowing to solve the two differential
equations with the Hunter's method \cite{HUN1}, as shown in the next section.
The solution of the field equations is obtained considering equations in vacuum,
in which case, the energy-momentum tensor vanishes and the content of matter is
expressed as a boundary condition on the fields in the disk, as shown below.
Finally, in section \ref{s_apli_mod} the first axially symmetric models in the
1PN approximation are presented.

\section{Formulation of the method} \label{s_form_met}

For thin disks of finite radius the components of the energy-momentum tensor can
be written as
\begin{eqnarray}
\stackrel{0 \: \: \: \: }{T^{00}} &=& c^2 {\Sigma}(R) \delta(z),
\label{T000-Sigma}\\
\stackrel{2 \: \: \: \: }{T^{00}} + \stackrel{2 \: \: \: \: }{T^{ii}} &=&
{\sigma}(R) \delta(z), \label{Tii02-sigma}
\end{eqnarray}
for $0 \leq R \leq a$, where $\delta$ is the Dirac delta function, and being
zero for $R>a$. Therefore, the field equations reduce to two Laplace equations
for the fields $\phi$ and $\psi$. It is demanded that the fields are even
functions of $z$
\begin{equation}
\phi(R,z)= \phi(R,-z), \qquad \psi(R,z)= \psi(R,-z),\label{reflection1}
\end{equation}
and therefore, that the first derivatives with respect to $z$ are odd functions
of $z$.

Using Gauss' theorem with (\ref{T000-Sigma}) and (\ref{Tii02-sigma}) gives,
\begin{eqnarray}
&& {\Sigma}(R) = \frac{1}{2 \pi G}\left(\frac{\partial \phi}{\partial
z}\right)_{z=0^{+}}, \label{Sigma-phi}\\
&& {\sigma}(R) = \frac{1}{2 \pi G}\left(\frac{\partial \psi}{\partial
z}\right)_{z=0^{+}}. \label{sigma-psi}
\end{eqnarray}
The problem is defined with the following boundary conditions: fields vanish at
infinity, and at $z=0$ its derivatives depend on the energy-momentum tensor
components according to (\ref{Sigma-phi}) and (\ref{sigma-psi}) for $0\leq R\leq
a$ and vanishing for $R>a$. Applying Hunter's method to each of the 1PN field
equations, one can obtain exact analytical expressions for the potentials. The
Hunter's method consists on obtaining solutions of the Laplace equation in
oblate spheroidal coordinates. The oblate coordinates are related to the
cylindrical by
\begin{eqnarray}
&& R = a \sqrt{(1+\xi^{2}) (1-\eta^{2})}, \label{oblatas_R} \\
&& z = a \xi \eta,\label{oblatas_z}
\end{eqnarray}
where $0 \leq \xi<\infty$ and $-1\leq\eta\leq 1$. The disk is placed at $\xi =
0$, where, $\eta^{2}=1-R^{2}/a^{2}$.

Following Hunter \cite{HUN1}, the general solution of each Laplace equation
satisfying previous boundary conditions can be written as
\begin{equation}
\phi(\xi,\eta) = - \sum_{n=0}^{\infty} A_{2n} q_{2n}(\xi) P_{2n}(\eta),
\label{phi1}
\end{equation}
for the $\phi$ potential, and
\begin{equation}
\psi(\xi,\eta) = - \sum_{n=0}^{\infty} B_{2n} q_{2n}(\xi) P_{2n}(\eta),
\label{psi1}
\end{equation}
for the $\psi$ potential, where $A_{2n}$ and $B_{2n}$ are the constants required
for each model, $P_{2n}(\eta)$ are the Legendre polynomials and
$q_{2n}(\xi)=i^{2n+1}Q_{2n}(i\xi)$ are Legendre functions of second kind. Note
that when using classical models, the expression for $\phi$ can be written
taking constants of the form
\begin{equation}
A_{2n} = C_{2n}+D_{2n}/c^{2},\label{constantesA}
\end{equation}
where the $C_{2n}$ constants define the Newtonian potential $\phi_N$ and the
constants $D_{2n}$ define the correction $\phi_{PN}$. So that taking the limit
$c\rightarrow\infty$, $\phi=\phi_{N}+\phi_{PN}$ is reduced to the Newtonian part
only. The corresponding expressions for $\Sigma$ and $\sigma$ in oblate
coordinates are
\begin{eqnarray}
\Sigma = \frac{1}{2\pi a G \eta}\sum_{n = 0}^{\infty} A_{2n} (2n+1) q_{2n+1}(0)
P_{2n}(\eta),&& \label{Sigma1}\\
&&\nonumber\\
\sigma = \frac{1}{2\pi a G \eta}\sum_{n = 0}^{\infty} B_{2n} (2n+1) q_{2n+1}(0)
P_{2n}(\eta).&& \label{sigma1}
\end{eqnarray}
Accordingly, $\Sigma$ can also be writen as the sum of a Newtonian part and a
post-Newtonian correction: $\Sigma = \Sigma_{N} + \Sigma_{PN}$.

In order to obtain self-consistent models of axially symmetric thin disks in
equilibrium, it is used a DF of the form $F = f(R,v_{R},v_{\varphi}) \delta(z)
\delta(v_{z})$, which is zero for $R>a$. The DF reproduces $\Sigma$ and $\sigma$
through the equations
\begin{eqnarray}
{\Sigma}(R) &=& \int \int f(R,v_{R},v_{\varphi}) d v_{R}dv_{\varphi},
\label{relacion-f-Sigma}\\
{\sigma}(R) &=& 4 \int \int E f(R,v_{R},v_{\varphi}) d v_{R}dv_{\varphi} -
2\phi_N \Sigma_N, \label{relacion-f-sigma}
\end{eqnarray}
where $E$ is the integral of motion defined by (\ref{ec_ene}). As in the
Newtonian case, one could define a relative energy and a relative potential as
\begin{eqnarray}
\varepsilon &=& -E + \Phi_0, \label{ec_ene_lz_pn}\\
\Psi &=&-\Phi + \Phi_0, \label{ec_pot efec_lz_pn}
\end{eqnarray}
where $\Phi_0$ is a constant that is chosen so that $\varepsilon$ and $\Psi$ are
always positive, i.e., such that $f>0$ for $0<\varepsilon\leq\Psi$.

The method to obtain self-consistent post-Newtonian models is to take
(\ref{relacion-f-Sigma}) with (\ref{Sigma1}) as the first fundamental equation
and (\ref{relacion-f-sigma}) with (\ref{sigma1}) as the second fundamental
equation. In accordance with the order of magnitude of the approach, the second
fundamental equation is solved using the Newtonian terms and then the first
fundamental equation is solved using the terms up to 1PN order. Finally, note
that for the second fundamental equation, the constants can be obtained directly
by
\begin{equation}
B_{2i}=\frac{4i+1}{4i+2}\frac{2\pi
aG}{q_{2i+1}(0)}\int_{-1}^{1}P_{2i}(\eta)\eta\sigma\mathrm{d}\eta,
\label{alphas_despejadas}
\end{equation}
where the integral depends on the particular model. On the other hand, in the
first fundamental equation, the constants also appear on the right side of the
equation (hence, it is necessary to obtain the constants differently for each
model) and, in general, we can not use an explicit expression as done above.
However, it is found that for all models treated it is finally possible to
obtain expressions analog to (\ref{alphas_despejadas}) for the first fundamental
equation, see (\ref{D0}) - (\ref{D4}) and (\ref{consphi1PN-kalnajs2}) -
(\ref{consphi1PN-kalnajs2b}). The DF in the 1PN approximation presents the same
functional dependence on the integrals of motion that in the Newtonian models,
that is, one uses the same DF but now with the 1PN energy and angular momentum.
Of course, this is for the sake of correspondence with the Newtonian limit. Note
that the correspondence principle must be satisfied by both the DF and its
integrals.

It is possible to obtain alternative expressions for $\Sigma$ and $\sigma$
considering in particular the case where the DF depends on a linear combination
of energy and angular momentum called Jacobi's integral, $J=\Omega L_{z}-E$.
Jacobi's integral is interpreted classically as the energy measured from a frame
of reference rotating with angular speed $\Omega$ \cite{BT}. In terms of the
relative energy (\ref{ec_ene_lz_pn}), we can write Jacobi's integral as
\begin{equation}
J = \varepsilon + \Omega L_z - \Psi_e(0), \label{integral-jacobi}
\end{equation}
where $\Psi_e(0)$ is the 1PN relative-effective potential evaluated in $\eta=0$,
defined by
\begin{equation}
\Psi_e = \Psi + \begin{matrix}\frac{1}{2}\end{matrix} \Omega^{2}R^{2}
\left(1-2\phi/c^{2}\right). \label{pot_rel}
\end{equation}
The Jacobi's integral takes values between zero and $J_{max}$, with
\begin{equation}
J_{max}  = \Psi - \frac{1}{2} \Omega^{2} a^{2} \eta^{2} \left(1 -
\frac{2\phi}{c^{2}} \right) - \frac{\Omega^{2} a^{2} \phi}{c^{2}}. \label{Jmax}
\end{equation}
Now, given that $2\pi dJ=dv_{R}dv_{\varphi}$, the relation
(\ref{relacion-f-Sigma}) can be written as
\begin{equation}
{\Sigma} = 2 \pi \int_{0}^{J_{max}} f(J) d J. \label{relacion-f-Sigma-J}
\end{equation}
Then, by using the expression
\begin{equation}
\int\int v_{\varphi} f dv_{R} dv_{\varphi} = 2 \pi \langle v_{\varphi} \rangle
\int_{0}^{J_{max}} f(J) d J,
\end{equation}
the second fundamental equation can be written as
\begin{equation}
{\sigma} = 2(2\Phi_{0}-\Omega^{2}a^{2}-\phi_N+2a\Omega \sqrt{1-\eta^{2}}\langle
v_{\varphi}\rangle)\Sigma_N - 8\pi\int_{0}^{J_{max}}J f(J) d J ,
\label{relacion-f-sigma-J}
\end{equation}
with $\Sigma_N$ given by (\ref{Sigma1}) with $A_{2n} = C_{2n}$.

Finally, the circular speed necessary for the rotation curve can be obtained
considering (\ref{ecmovimiento1PN}) for circular orbits. In that case, the
circular speed is equal to $v_{\phi}$ and perpendicular to the gradient of the
field $\phi$. Therefore, the 1PN equation of motion (\ref{ecmovimiento1PN})
reduces to
\begin{equation}
\frac{v_{\varphi}^{2}}{R} \left[ 1 + \frac{R}{c^{2}} \frac{\partial
\phi}{\partial R} \right]_{z=0} = \frac{\partial}{\partial R} \left[ \phi +
\frac{2 \phi^{2} + \psi}{c^{2}} \right]_{z=0}.
\end{equation}
Then, in accordance with the 1PN order of approximation, the expression for the
circular speed is
\begin{equation}
v_{\varphi} = \left. \sqrt{R \frac{\partial \phi}{\partial R} \left(1 + \frac{4
\phi}{c^{2}} - \frac{R}{c^{2}} \frac{\partial \phi}{\partial R} \right) +
\frac{R}{c^{2}} \frac{\partial \psi}{\partial R}} \right \rfloor_{z=0}.
\label{circular-velocity}
\end{equation}
Note that in the limit $c \rightarrow \infty$, the previous expression reduces
to that of Newtonian theory $v_{\varphi}=\sqrt{R\partial\phi_N/\partial R}$.

\section{Application to the Generalized Kalnajs Disks} \label{s_apli_mod}

We will now apply the previous formalism to the family of Generalized Kalnajs
Disks (GKD) introduced by Gonz\'alez and Reina in \cite{GR}. This family is
characterized by mass densities of the form
\begin{equation}\label{dens-kalnajs}
\Sigma_{N}^{(m)} = \frac{3M}{2\pi a^{2}} \eta^{2m-1},
\end{equation}
where the index $m$ of the model is any positive integer. The potential
$\phi_N^{(m)}$ is given by (\ref{phi1}) by taking the Newtonian limit in
(\ref{constantesA}) and using
\begin{equation}\label{C2n}
C_{2n}^{(m)} = \frac{M G\pi^{1/2} (4n+1) (2m+1)!}{a2^{2m+1} (2n+1) (m - n)!
\Gamma(m + n + \frac{3}{2} ) q_{2n+1}(0)}
\end{equation}
for $n \leq m$, and $C_{2n}^{(m)} = 0$ for $n > m$.

\subsection{The 1PN model for the $m = 1$ GKD}

The first disk of the family, the disk with $m = 1$, is the well known Kalnajs
disk \cite{KAL1}, with the mass distribution
\begin{equation}\label{dens-kalnajs1}
\Sigma_{N}^{(1)} = \frac{3M \eta}{2\pi a^{2}}.
\end{equation}
Then, as was shown by Kalnajs \cite{KAL2}, the DF depends on the Jacobi's
integral as
\begin{equation}\label{DF-kalnajs1}
f(J) = \frac{3M}{4\pi^{2}
a^{3}}\left[2(\Omega_{0}^{2}-\Omega^{2})J\right]^{-1/2}
\end{equation}
where $\Omega_0^{2}=3\pi GM/4a^{3}$. Now, for the Kalnajs disk it holds that
$\langle v_{\varphi}\rangle=\Omega R$, so that the disk spins like a rigid body.

Inserting expressions (\ref{dens-kalnajs1}) and (\ref{DF-kalnajs1}) into the
second fundamental equation (\ref{relacion-f-sigma-J}), and integrating
the DF, one easily obtains the following system of equations
\begin{eqnarray}
B_{0} - B_{2} + B_{4} &=& 0, \label{eq4Kalnajs1PN}\\
3 B_{2} - 10 B_{4} &=& 6 M G a \Omega^{2} - \frac{27 \pi G^{2} M^{2}}{4 a^{2}},
\label{eq5Kalnajs1PN}\\
B_{4} & = & \frac{9 \pi G^{2} M^{2}}{140 a^{2}} - 30 G M \Omega^{2}
a. \label{eq6Kalnajs1PN}
\end{eqnarray}
Likewise, for the first fundamental equation we have
\begin{eqnarray}
D_{0} - D_{2} + D_{4} &=& \frac{9 \pi G^{2} M^{2} \Omega^{2}}{8 a^{2}
(\Omega_{o}^{2} - \Omega^{2})}, \label{eq1Kalnajs1PN}\\
\gamma D_{2} + \vartheta D_{4} &=& B_{2} - \frac{15}{8} B_{4},
\label{eq2Kalnajs1PN}\\
\chi D_{4} &=& \frac{35}{16} B_{4} - \frac{3 \pi G^{2} M^{2}}{4a^{2}} - G M
\Omega^{2} a, \label{eq3Kalnajs1PN}
\end{eqnarray}
where
\begin{eqnarray}
\gamma &=& \frac{24 a^{3}(\Omega_{o}^{2} - \Omega^{2}) - 9 \pi M G}{9 \pi M
G}, \label{gamma}\\
\vartheta &=& \frac{135 \pi M G/8 - 80 a^{3}(\Omega_{o}^{2} - \Omega^{2})}{9 \pi
M G}, \label{vartheta}\\
\chi &=& \frac{70 [128 a^{3} (\Omega_{o}^{2} - \Omega^{2}) - 27 \pi M G]}{864
\pi M G}.
\end{eqnarray}
Therefore, we have a system of linear equations with an upper triangular matrix
for the constants of each potential, $\psi$ and $\phi_{PN}$.

\begin{figure*}
$$\begin{array}{cc}
\epsfig{width=6.15cm,file=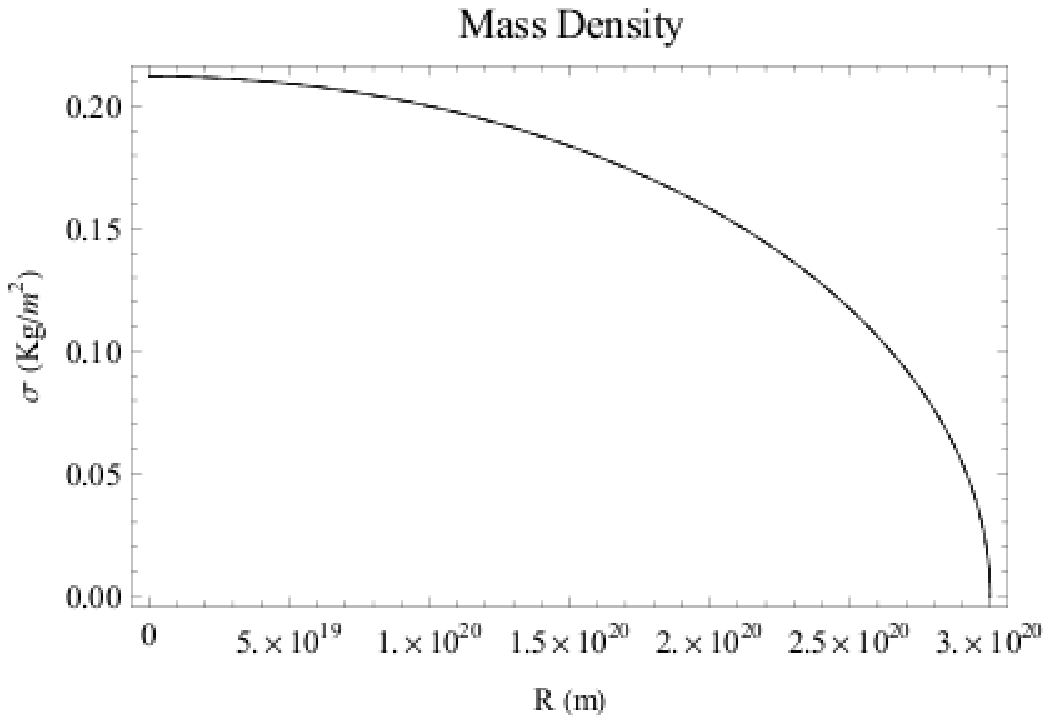} &
\epsfig{width=6.15cm,file=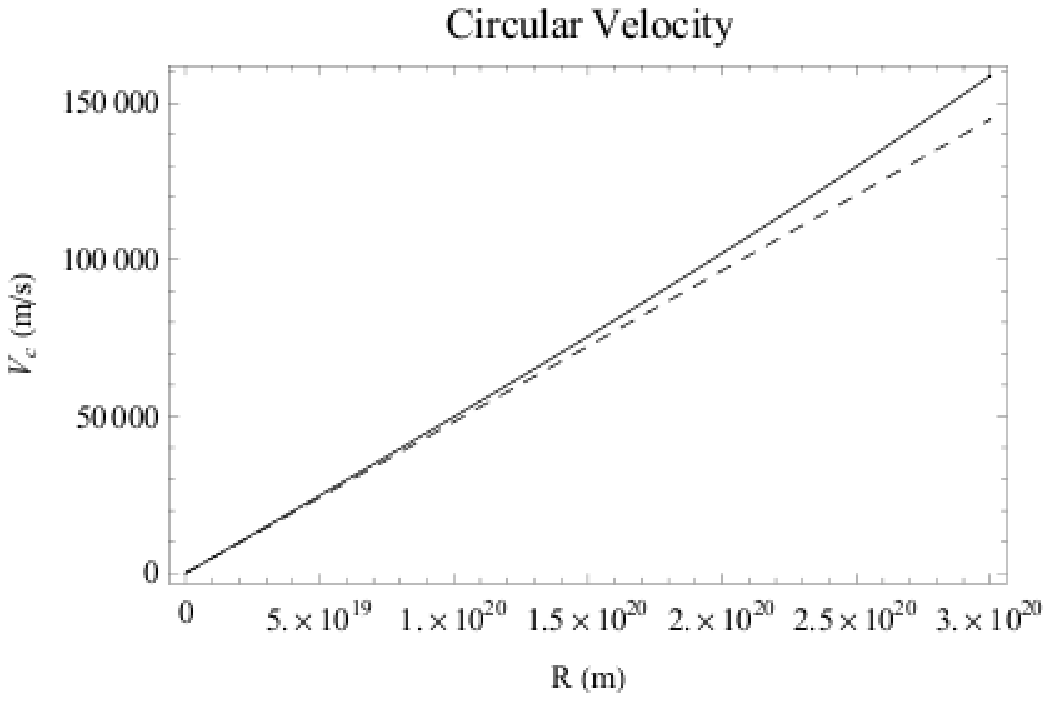} \\
\end{array}$$
\caption{First we plot the Newtonian and 1PN mass density for the $m = 1$ GKD.
The two curves seem to be superposed meaning that the mass correction is
negligible. Then we plot the Newtonian (dashed line) and 1PN (full line)
rotation curves for the same disk. The 1PN corrections are clearly notorious
reaching a maximum at the border of the disc. Parameter values are: $M=4\times
10^{40}kg$, $a=3\times 10^{20}m$, $\Omega=2\times 10^{-13}Hz$. } \label{gkd1}
\end{figure*}

Solving for the constants explicitly, we obtain
\begin{eqnarray}
B_{0} &=& 2 a G M (\pi-1) \Omega^2 - \begin{matrix}\frac{3 G^2 M^2 \pi  (15 \pi
- 1)}{20 a^2} \end{matrix}, \label{B0}\\
&&	\nonumber	\\
B_{2} &=& \begin{matrix} \frac{2}{7} a G M (7 \pi - 10) \Omega^2 - \frac{3 G^2
M^2 \pi  (21 \pi - 2)}{28 a^2} \end{matrix}, \label{B2}\\
&&	\nonumber	\\
B_{4} &=& \begin{matrix} \frac{18 G M}{35 a} \left(\frac{G M \pi }{8 a} -
\frac{5 a^2 \Omega^2}{3} \right) \end{matrix}, \label{B4}\\
&&	\nonumber	\\
D_{0} &=& \begin{matrix} - \frac{27 G^3 \pi ^2 M^3}{8 a^2 \left(a^3 \Omega^2 - 3
G M \pi \right)} \end{matrix} - \begin{matrix} \frac{47709 G^3 \pi^2 M^3}{448
a^2 \left(128 a^3 \Omega^2 - 357 G M \pi \right)} \end{matrix} \nonumber \\
&& - \begin{matrix} \frac{3 G^2 \pi (1 + 80 \pi) M^2}{320 a^2} \end{matrix} -
\begin{matrix} \frac{9 \left( 28 G^3 M^3 \pi^3 - 95 G^3 M^3 \pi^2 \right)}{28
a^2 \left(8 a^3 \Omega^2 - 21 G M \pi \right)} \end{matrix}, \label{D0} \\
&&	\nonumber	\\
D_{2} &=& - \begin{matrix} \frac{3 G^2 \pi (7\pi - 13) M^2}{28 a^2} \end{matrix}
- \begin{matrix} \frac{9 \left(28 G^3 M^3 \pi^3 - 95 G^3 M^3 \pi^2 \right) }{28
a^2 \left(8 a^3 \Omega^2 - 21 G M \pi \right)} \end{matrix}, \label{D2} \\
&&	\nonumber	\\
D_{4} &=& \begin{matrix} \frac{27 G^2 M^2 \pi \left(184 \Omega^2 a^3 + 39 G M
\pi \right)}{140 a^2 \left(128 a^3 \Omega^2 - 357 G M \pi \right)} \end{matrix},
\label{D4}
\end{eqnarray}
which defines the potential $\psi$, and the correction $\phi_{PN}$,
respectively. Hence we can get all the parameters of interest, in particular, we
observed that the mass correction, $\Sigma_{PN}$, is negligible compared with
the classical mass $\Sigma_N$. In contrast, the 1PN rotation curve obtained with
(\ref{circular-velocity}), is visibly separated from the Newtonian one from a
certain radius, the difference being maximum at the edge of the disk (see Fig.
\ref{gkd1}), where the difference between both reaches $10.3\%$
approximately. Those have been obtained with the typical values of a galaxy such
as the Milky Way. 

\subsection{The 1PN model for the $m = 2$ GKD} \label{ss_kalnajs2}

The procedure for the second disk is pretty much the same that for the first
one. Again, the DF has a simple form when written as a function of Jacobi's
integral,
\begin{equation}\label{DF-kalnajs2}
f(J) = \frac{2M}{\sqrt{3} a^{2}}\left(\frac{10 a^{3}}{G^{3} M^{3}\pi^{11}
J}\right)^{1/4},
\end{equation}
for a mean circular speed $\langle v_{\varphi}\rangle=\Omega R= \sqrt{15 \pi G
M/32 a^{3}}R$. This DF could be easily integrated to self-consistently obtain
the mass density of the model, which also can be obtained from
(\ref{dens-kalnajs}) with $m=2$,
\begin{equation}\label{dens-kalnajs2}
\Sigma_{N}^{(2)} = \frac{5M}{2 \pi a^{2}} \left(1 - \frac{R^{2}}{a^{2}}
\right)^{3/2} = \frac{5M \eta^{3}}{2\pi a^{2}}.
\end{equation}
the associated gravitational potential is given by the Newtonian limit $\phi_N$
of (\ref{phi1}) with (\ref{C2n}) and $m=2$.

Again, we replace the Newtonian terms into the second fundamental equation
(\ref{relacion-f-sigma-J}) to obtain:
\begin{eqnarray}
\sum_{n = 0}^{4} B_{2n} (2n+1) q_{2n+1}(0) P_{2n}(\eta) = \sum_{n=0}^{4}
\tilde{C}_{2i} \eta^{2i}, \label{ecFundamental2-Kalnajs2}
\end{eqnarray}
where $\tilde{C}_{0} = \tilde{C}_{2} = 0$ and
\begin{eqnarray}
 \tilde{C}_{4} &=& (25\pi^{2}10! G^{2}M^{2})/(160 a^{2}), \\
 &&	\nonumber \\
 \tilde{C}_{6} &=&  - (75\pi^{2}10! G^{2}M^{2})/(160 a^{2}) \\
 &&	\nonumber \\
 \tilde{C}_{8} &=& (75\pi^{2}10! G^{2}M^{2})/(2240 a^{2}).
\end{eqnarray}
Multiplying (\ref{ecFundamental2-Kalnajs2}) with a Legendre polynomial,
integrating with respect to $\eta$, and using the orthogonality properties of
the Legendre polynomials we get
\begin{equation}\label{conspsi-kalnajs2}
B_{2n} = \sum_{i=0}^{4} \frac{\sqrt{\pi} \tilde{C}_{2i} 2^{-2i-1} (4n+1)
\Gamma(2n+1)}{q_{2n+1}(0) (2n+1) \Gamma(i-n+1) \Gamma(i+n+3/2)}.
\end{equation}
We also could have used the expression (\ref{alphas_despejadas}).

\begin{figure*}
$$\begin{array}{cc}
\epsfig{width=6.15cm,file=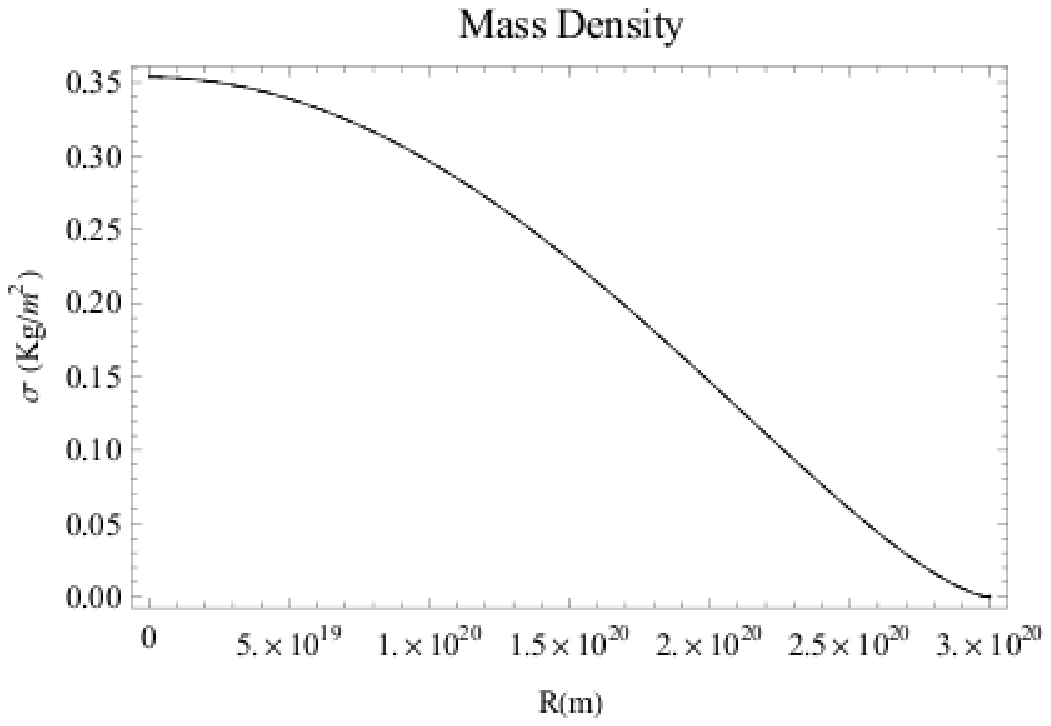} &
\epsfig{width=6.15cm,file=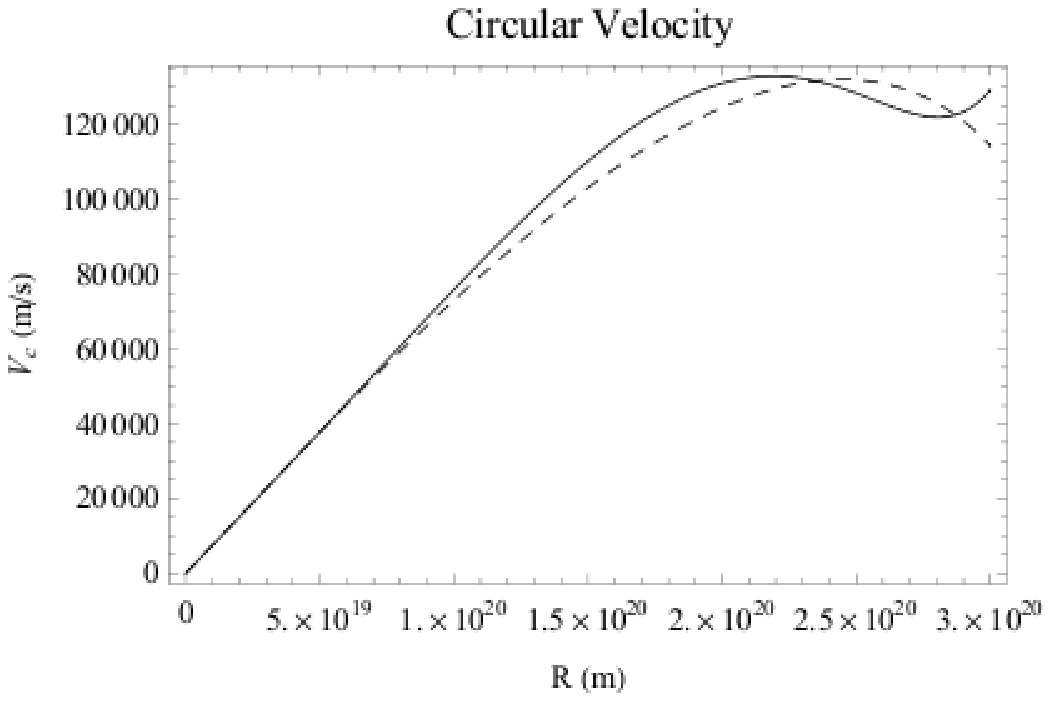} \\
\end{array}$$
\caption{We plot the Newtonian and 1PN mass density, as well as the Newtonian
(dashed line) and 1PN (full line) rotation curves for the $m = 2$ GKD, with
parameter values: $M=4\times 10^{40}kg$, $a=3\times 10^{20}m$, $\Omega=2\times
10^{-13}Hz$. As we can see, the two graphics of mass density appear to be
superposed, while that the 1PN rotation curve behaves completely different to
the Newtonian one.}\label{gkd2}
\end{figure*}

After rearranging the terms, the first fundamental equation can be written as
\begin{equation}
\sum_{n = 0}^{4} \{ D_{2n} [ \vartheta_{2n} P_{2n}(\eta) + q_{2n}(0)
P_{2n}(0) ] - B_{2n} q_{2n}(0) P_{2n}(\eta) - \hat{C}_{2n} \eta^{2n}
\} = 0, \label{ecFundamental1-Kalnajs2}
\end{equation}
where,
\begin{eqnarray}
\vartheta_{2n} & = & \pi(2j+1)/(32 a^{2} q_{2n+1}(0) - q_{2n}(0)), \\
 &&	\nonumber \\
\hat{C}_{0} & = & (675\pi^{2} G^{2}M^{2})/(4096 a^{2})+\psi(0,0), \\
 &&	\nonumber \\
\hat{C}_{2} & = & -(1575\pi^{2} G^{2}M^{2})/(4096 a^{2}), \\
 &&	\nonumber \\
 \hat{C}_{4} & = & -(1125\pi^{2} G^{2}M^{2})/(2048 a^{2}), \\
 &&	\nonumber \\
 \hat{C}_{6} & =&  - (2025\pi^{2} G^{2}M^{2})/(4096 a^{2}), \\
 &&	\nonumber \\
 \hat{C}_{8} & = & (2025\pi^{2} G^{2}M^{2})/(8192 a^{2}).
\end{eqnarray}
Integrating and using orthogonality relations, the constants $D_{2n}$ are given
by
\begin{equation}
D_{2n} = \sum_{i=0}^{4} \frac{\sqrt{\pi} \hat{C}_{2i} 2^{-2i-1} (4n + 1)
\Gamma(2j + 1)}{\vartheta_{2n} (2n + 1) \Gamma(i - n + 1) \Gamma(i + n + 3/2)} +
\frac{B_{2n} q_{2n}(0)}{\vartheta_{2n}}, \label{consphi1PN-kalnajs2}
\end{equation}
for $n>0$, and
\begin{equation}
D_{0} = \frac{1}{\vartheta_{0} + \pi/2} \left[\sum_{i=0}^{4} \frac{\sqrt{\pi}
\hat{C}_{2i} 2^{- 2i - 1} \Gamma(2j + 1)}{\Gamma(i + 1) \Gamma(i + 3/2)}  -
\sum_{i=1}^{4} D_{2i} q_{2i}(0) P_{2i}(0) \right], \label{consphi1PN-kalnajs2b}
\end{equation}
which defines the correction $\phi_{PN}$ to the Newtonian potential. The mass
densities and rotation curves are plotted at Figure \ref{gkd2}, note that the
rotation curve is not only cuantitative, but cualitatively different, presenting
its maximum value more closely to the center of the disk and going upwards at
the border. This is due to the fact that the 1PN curve includes more terms than
the Newtonian one.

\begin{thebibliography}{9999}

\bibitem{LB} D. Lynden-Bell, MNRAS 123, 447, (1962).

\bibitem{mestel} L. Mestel, MNRAS 126, 553 (1963).

\bibitem{T1} A. Toomre, ApJ 138, 385 (1963).

\bibitem{T2} A. Toomre, ApJ 139, 1217, (1964).

\bibitem{HunterToomre} C. Hunter and A. Toomre, ApJ 155, 747 (1969).

\bibitem{KAL1} A. J. Kalnajs, ApJ 175, 63, (1972).

\bibitem{jiang00} Z. Jiang, MNRAS 319, 1067, (2000).

\bibitem{jiangmos02} Z. Jiang and  D. Moss, MNRAS 331, 117, (2002).

\bibitem{GR} G. A. Gonz\'alez and J. I. Reina, MNRAS 371, 1873 (2006).

\bibitem{jiang} Z. Jiang and L. Ossipkov, MNRAS 379,1133, (2007).

\bibitem{PRG} J. F. Pedraza, J. Ramos-Caro and G. A. Gonz\'alez, MNRAS 390,
1587, (2008).

\bibitem{Lemos1} J. P. S. Lemos and  P. S. Letelier, Class. Quantum Grav. 10,
L75, (1993).

\bibitem{Lemos2} J. P. S. Lemos and  P. S. Letelier, Phys. Rev. D 49, 5135,
(1994).

\bibitem{Lemos3} J. P. S. Lemos and  P. S. Letelier, Int. J. Modern Phys. D 5,
53, (1996).

\bibitem{gonzalez-letelier} G. A. Gonz\'alez and P. S. Letelier, Class. Quantum
Grav. 16, 479, (1999).

\bibitem{gonzalez-letelier2} G. A. Gonz\'alez and P. S. Letelier, Phys. Rev. D
62, 064025, (2000).

\bibitem{semerak-zacek1} O. Semer\'ak and M. Z\'acek, Class. Quantum Grav. 17,
1613, (2000).

\bibitem{semerak-zacek2} O. Semer\'ak and M. Z\'acek, Class. Quantum Grav. 19,
3829, (2002).

\bibitem{semerak-zacek3} M. Z\'acek and O. Semer\'ak, Czech. J. Phys. 52, 19,
(2002).

\bibitem{Rezania} V. Rezania and Y. Sobouti, Astron. Astrophys. 354, 1110,
(2000).

\bibitem{BT} J. Binney and S. Tremaine, {\it Galactic Dynamics}, 2nd Ed.
Princeton University Press, Princeton, N. J., (2008).

\bibitem{HUN1} C. Hunter, MNRAS 126, 299, (1963).

\bibitem{KAL2} A. J. Kalnajs, ApJ 205, 751, (1976).

\end{thebibliography}
\end{document}